\documentclass[num-refs]{wiley-article}

\usepackage{siunitx}

\papertype{Original Article}

\paperfield{Journal Section}

\title{Anti-counterfeiting tags with camouflaged QR codes on nanocavities, using polymer-dispersed-liquid-crystals}

\author[1,2\authfn{1}]{Giuseppe Nicoletta}
\author[2\authfn{1}]{Mauro Daniel Luigi Bruno}
\author[3,4]{Peng Yu}
\author[3,4]{Zhiming Wang}
\author[1,2]{Maria Penelope De Santo}
\author[1,2,3]{Roberto Caputo}
\author[2]{Antonio Ferraro}

\contrib[\authfn{1}]{Equally contributing authors.}

\affil[1]{University of Calabria, Physics Department, \\87036 Rende (CS), Italy}
\affil[2]{Consiglio Nazionale delle Ricerche - Istituto di Nanotecnologia CNR-Nanotec, Rende (CS), 87036 Italy}
\affil[3]{Institute of Fundamental and Frontier Sciences, University of Electronic Science and Technology of China, Chengdu 610054, P.R. China}
\affil[4]{Shimmer Center, Tianfu Jiangxi Laboratory, Chengdu 641419, P. R. China}

\corraddress{Giuseppe Nicoletta, Mauro Daniel Luigi Bruno, Roberto Caputo}
\corremail{giuseppe.nicoletta@unical.it; bruno.mauro89@gmail.com; roberto.caputo@unical.it}

\presentadd[\authfn{2}]{Department, Institution, City, State or Province, Postal Code, Country}

\fundinginfo{Funder One, Funder One Department, Grant/Award Number: 123456, 123457 and 123458; Funder Two, Funder Two Department, Grant/Award Number: 123459}

\runningauthor{Nicoletta et al.}

\begin{document}

\begin{frontmatter}
\maketitle

\begin{abstract}
Counterfeiting poses an evergrowing challenge, driving the need for innovative and sophisticated anti-counterfeiting strategies and technologies. Many solutions focus on tags characterized by optical features that are partially or completely camouflaged to the human eye, thus discouraging scammers. In this paper, a QR code is laser printed on a thin plastic foil previously coated by a specific nanocavity consisting of a metal/insulator/metal/insulator (MIMI) multilayer. This metamaterial possesses unique features in terms of light transmission that are due to the specific design. 
A thin layer of polymer dispersed liquid crystals, fabricated incorporating specific nematic liquid crystals in a polymer matrix, is able to camouflage the QR code that becomes, then, readable only under specific thermal conditions. Three anti-counterfeiting tags were fabricated, each using a distinct LC with its own nematic-isotropic transition temperature. The peculiar combination of the unique optical properties of nematic liquid crystals and optical nanocavities results in the creation of a novel type of tags showing two different encoding levels. 
Stress tests including water immersion, bending test, and prolonged heating have been performed ensuring the long-term stability of the tags. The realized two security-level anti-counterfeiting tags are cost-effective, straightforward to manufacture and, thanks to their flexibility, can be easily integrated into packaging and products.

\keywords{anti-counterfeiting, nanocavity, polymer dispersed liquid crystal (PDLC)}
\end{abstract}
\end{frontmatter}

\section{Introduction}
Over the recent decades, the rise in counterfeit products has prompted the scientific community to invest substantial resources in developing solutions to ensure product authenticity. Counterfeiting not only impacts companies at financial level, but also poses serious risks to human health since the phenomenon affects not only luxury goods and clothing but also food and medicines. The causes often include ineffective solutions, such as simple QR codes, barcodes, and holograms, which can be easily cloned due to their deterministic production, or costly systems that may exceed the value of the products \cite{lehtonen2007features, javidi2016roadmap}. To address these challenges, scientific research is increasingly focused on developing anti-counterfeiting tags based on the physical unclonable functions (PUFs) paradigm. Indeed, PUFs have emerged as a promising technology that utilizes the intrinsic physical properties of materials to generate unique, hard-to-replicate identifiers. The uniqueness of a PUF key lies in the non-deterministic features arising during the production process. Generally, a PUF produces an output (response) following an input (challenge), exploring therefore the challenge-response pair scheme \cite{arppe2017physical, herder2014physical, gao2020physical}. 
Such tags then work as ”fingerprints”, offering a high level of security while maintaining accessibility and affordability \cite{wu2022high,wang2022physical,mispan2021physical}.
In this perspective, different materials and methods are used to realize PUF key-based tags, at micro- and even nano-scale, as carbon nanotube field-effect transistors \cite{gao2016emerging}. 
Soft matter, as liquid crystals (LCs) and polymers, is a further excellent candidate for the creation of PUFs. As an example, block copolymers are used for their self-assembly properties \cite{kim2022nanoscale}. Liquid crystals are materials capable of self-assembling into complex photonic structures exhibiting peculiar optical properties. The most interesting optical properties include birefringence \cite{kim2023voxelated}, Bragg reflection \cite{lenzini2017security}, and the ability to modulate these properties by controlling external stimuli such as radiation, electric and magnetic fields, and heat. In 2024 Nocentini et al. developed reconfigurable structures using light-transformable polymers, thus enabling the possibility to reversibly reconfigure the physical structure of the unclonable function \cite{nocentini2024all}.
Another key issue involves the authentication process: many proposed PUFs require complex laboratory equipment, such as Scanning Electron Microscopy \cite{lin2024structural,rezaei2023tri}, Atomic Forced Microscopy \cite{smith2018microlithography,gates2005new}, Raman spectrometer, and others for validation \cite{kayaci2022organic,silverio2022functional}. 
These challenges underscore the need for low-cost, easily verifiable, smart tags using both affordable materials and simple authentication methods accessible through devices like smartphones. 
During the last years, we developed different PUF key based anti-counterfeiting tags exploiting soft matter, as liquid crystals and polymers, which can be authenticated using a smartphone \cite{bruno2023cholesteric,bruno2024flexible}.

An example is reported in 2022 in which Ferraro et al. \cite{ferraro2023hybrid} proposed a multilevel label based on a QR code printed on office paper, camouflaged with an optical cavity. The QR code becomes readable when back-illuminated, exploiting the optical properties of the cavity and the random morphology of the paper as additional encoding levels. 
However, this tag functions exclusively with transmitted light, restricting its use to transparent items.
 
To overcome this limitation, here we propose a multilevel PUF-inspired anti-counterfeiting tag featuring a QR code camouflaged with polymer dispersed liquid crystal (PDLC), which can be authenticated through standard procedures. The PDLCs were fabricated with three distinct liquid crystals, each with isotropic transition temperatures ranging from 35°C to 130°C. This temperature range ensures the QR code remains hidden, providing protection until the authentication process. When the PDLC is heated above its transition temperature, the QR code becomes visible.
Additionally, the proposed tag incorporates two security levels. The first level is a randomly generated QR code, readable with a smartphone camera once the PDLC becomes transparent. The second level leverages the unique optical properties of the nanocavity. In fact, the tag exhibits different colors depending if observed in reflection or transmission mode. 
Finally, the proposed tag is fabricated on a flexible substrate allowing it to adapt to both rigid and soft surfaces, while ensuring the quality and security of the product. 

\section{Experimental section}
\begin{figure} [h]
    \centering
    \includegraphics[width=0.5\columnwidth]{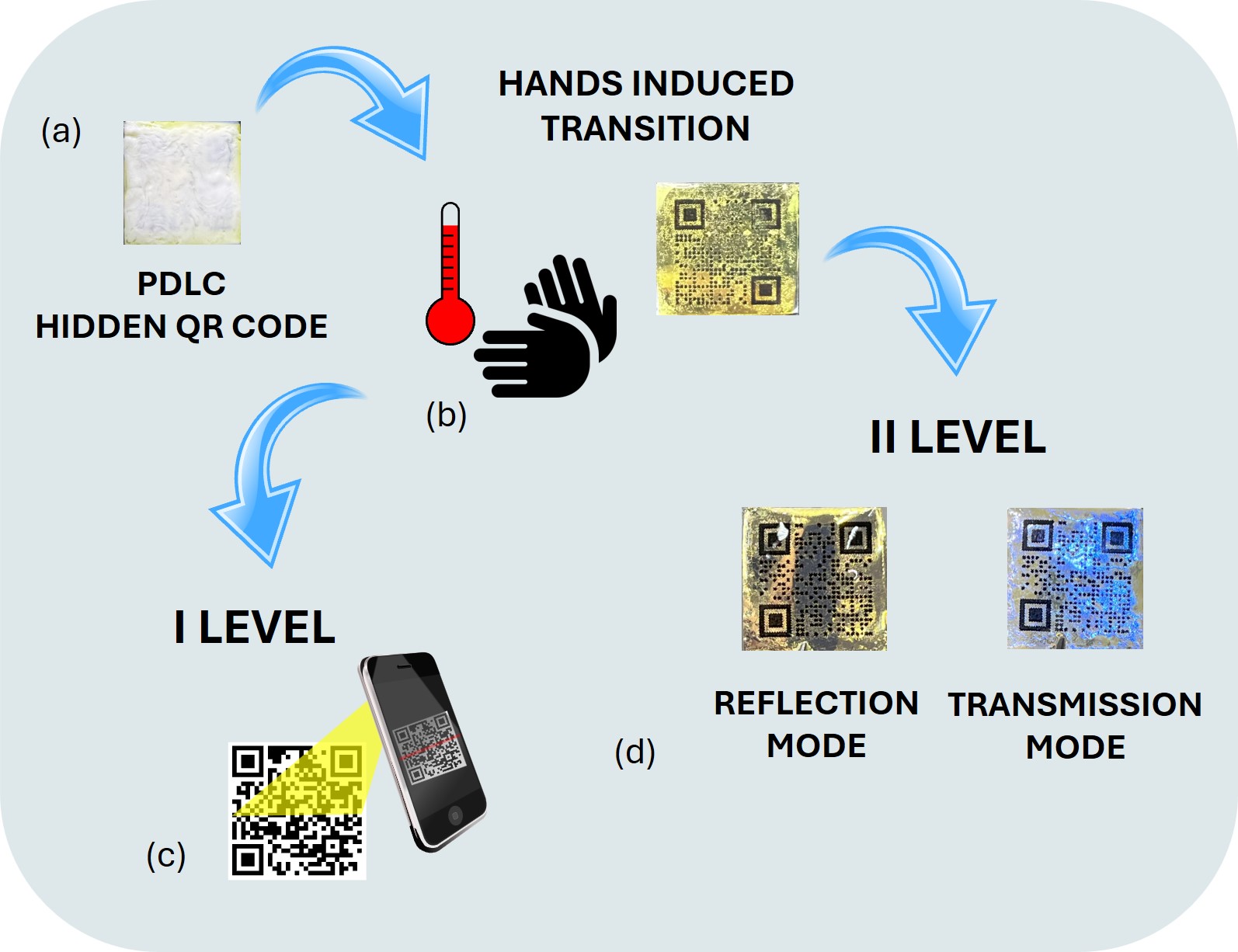}
    \caption{(a) Polymer Dispersed Liquid Crystal (PDLC) hidden QR code (b) visible when inducing nematic-isotropic transition rubbing hands. (c) After the transition temperature, the QR code, which represents the first encoding level, can be read using a smartphone. (d) Further, the second encoding level can be validated optically in reflection and transmission mode.}
    \label{Table}
\end{figure}

\begin{figure} [h]
    \centering
    \includegraphics[width=0.5\columnwidth]{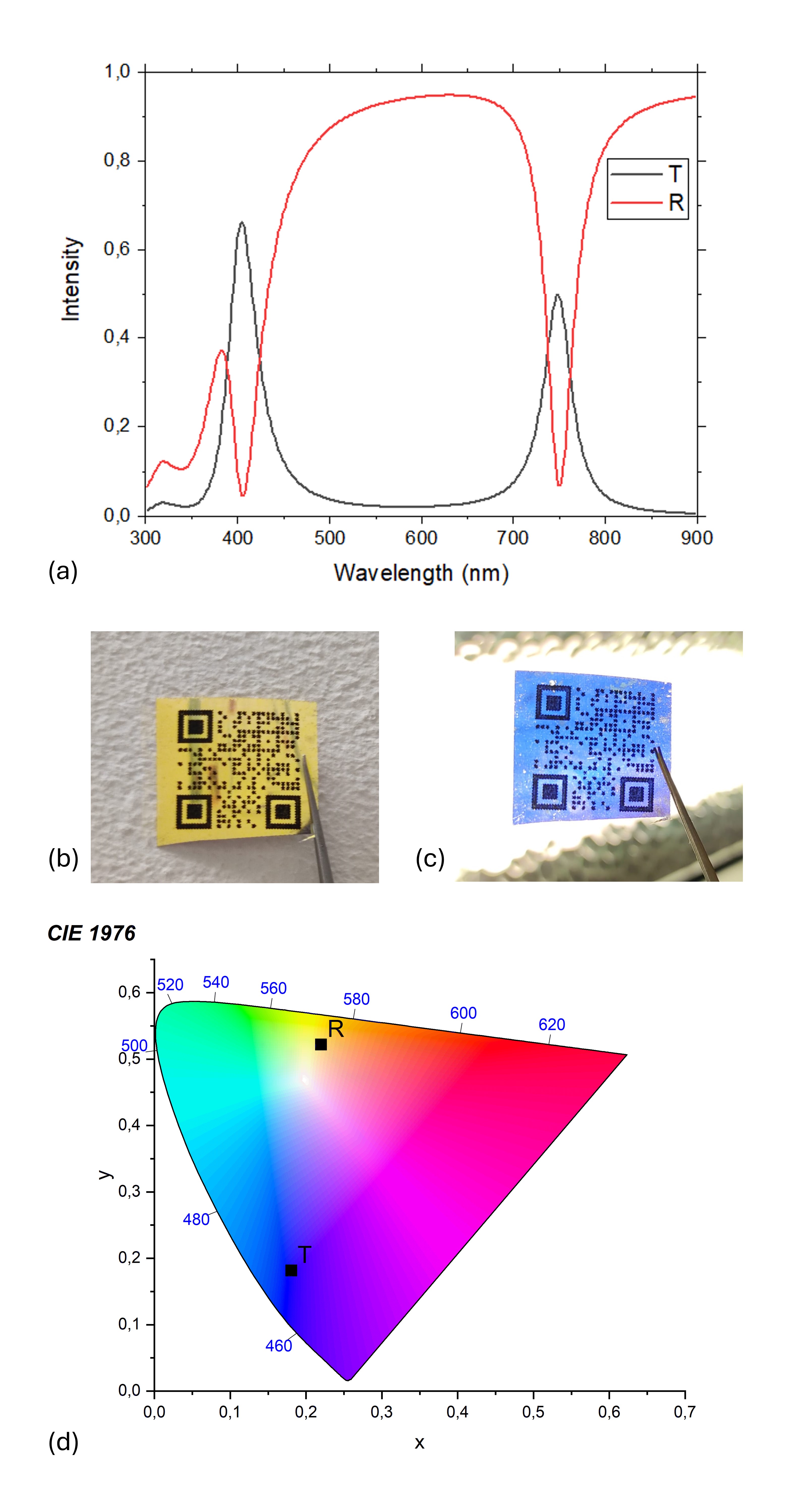}
    \caption{a) Numerical reflection (red line) and transmission  (black line) of the optical nanocavity composed by Ag (30nm), ZnO (150nm), Ag (30nm), ZnO(30 nm). b) and c) show a tag produced with sputtering on a flexible substrate in reflection and transmission mode respectively. (d) Related CIE 1976 of simulated transmission and reflection with the corresponding points.}
    \label{img1}
\end{figure}

The proposed anti-counterfeiting solution features a tag with a randomly generated QR code that remains invisible under normal conditions, such as room temperature (Figure \ref{Table}a). Depending on the isotropic transition temperature of the contained liquid crystal, the QR code becomes visible when the tag is heated using methods such as heat gun, hot plate, or simply by rubbing it between the hands (Figure \ref{Table}b). 
At this stage, the QR code can be validated (Figure \ref{Table}c), representing the first level of encoding. The second security level, relying on the peculiar optical properties of the nanocavity tag, can then be verified. The tag exhibits distinct colors under different illumination modes: gold in reflection (Figure \ref{Table}d, left) and blue in transmission (Figure \ref{Table}d, right).
The novelty of this anti-counterfeiting system lies in the unique combination of rigid and soft materials, which enables to obtain the distinct characteristics described above. In detail, the fabrication of the tag begins by depositing a multilayer metamaterial on a polyethylene (PET) foil, which serves as an optical nanocavity, The QR code is then printed onto this layer. Next, the entire assembly is sandwiched between two PET sheets. Finally, a PDLC layer is applied on top of the resulting structure. The metamaterial, which forms the core of the tag, is created using a metal-insulator-metal-insulator (MIMI) structure.
The tailored optical properties of the MIMI structure were determined through a transfer matrix method (TMM) simulation, considering Ag/ZnO/Ag/ZnO layers deposited on a flexible PET substrate. The assumed layer thicknesses (from the substrate) were 30 nm of silver, 150 nm of zinc oxide, 30 nm of silver, and a final 30 nm layer of zinc oxide. The resulting optical responses of the MIMI, shown in Figure \ref{img1} a, include reflection (red line) and transmission (black line)).
The chosen multilayer configuration presents features covering almost the entire UV-Vis spectral range with peaks/dips in transmission/reflection at $\lambda=405nm$ and $\lambda=748nm$. 
The physical deposition of the multilayer onto the PET plastic foil has been performed by DC sputtering (Kenosistec K300) at a vacuum condition of $7 * 10^{- 6}$ mbar. For the 30 nm silver layers, a DC power of 50 W for 105 s has been used. Whereas the 150 nm and 30 nm ZnO alternating layers were deposited using the RF cathode at a power of 80 W for 48 min 59 s and 9 min 41 s respectively.
QR codes linking to the website of the University of Calabria (for demonstration purposes), were laser printed on the previously realized multilayer structure. The QR code can be obviously randomly generated and represents the first level of security of our tag. 
In reflection, the fabricated tag displays a bright gold appearance (Figure \ref{img1} b), while in transmission, only the blue component is visible (Figure \ref{img1} c). This distinct optical behavior constitutes the tag's second security level. The experimentally obtained colors are in agreement with the numerical ones, which are reported in the 1976 chromaticity diagram (Figure \ref{img1}d).
To create a camouflaged QR code, three distinct mixtures of PMMA (12\% dissolved in 2-propanone ((CH3)2CO)) and liquid crystals were prepared, resulting in three unique PDLCs.
These mixtures were deposited onto three separate tags using a drop-casting technique. To protect the printed ink from the solvent used to dissolve the polymer, a transparent adhesive PET film was first applied to the tag. 

\begin{figure} [t]
    \centering
    \includegraphics[width=0.5\columnwidth]{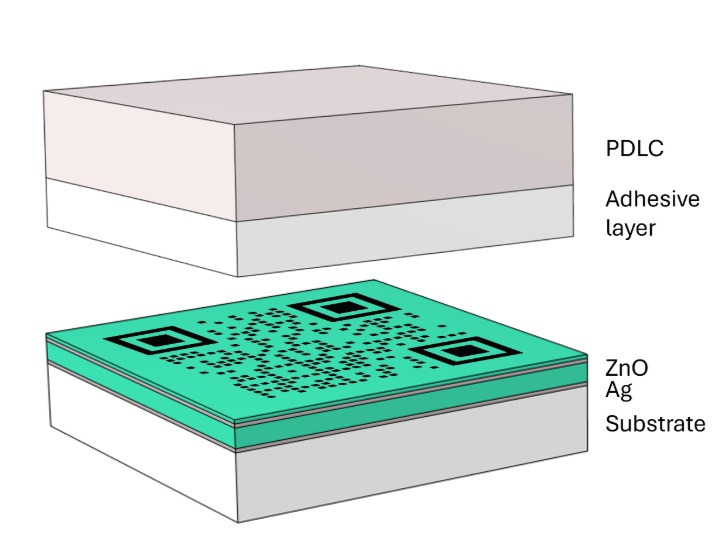}
    \caption{Schematic representation of the PDLC TAG. The PDLC and the adhesive layer are separated from the other parts of the TAG to show the position of the QR code.}
    \label{Img5}
\end{figure}

\begin{figure}[h]
    \centering
    \includegraphics[width=0.6\linewidth]{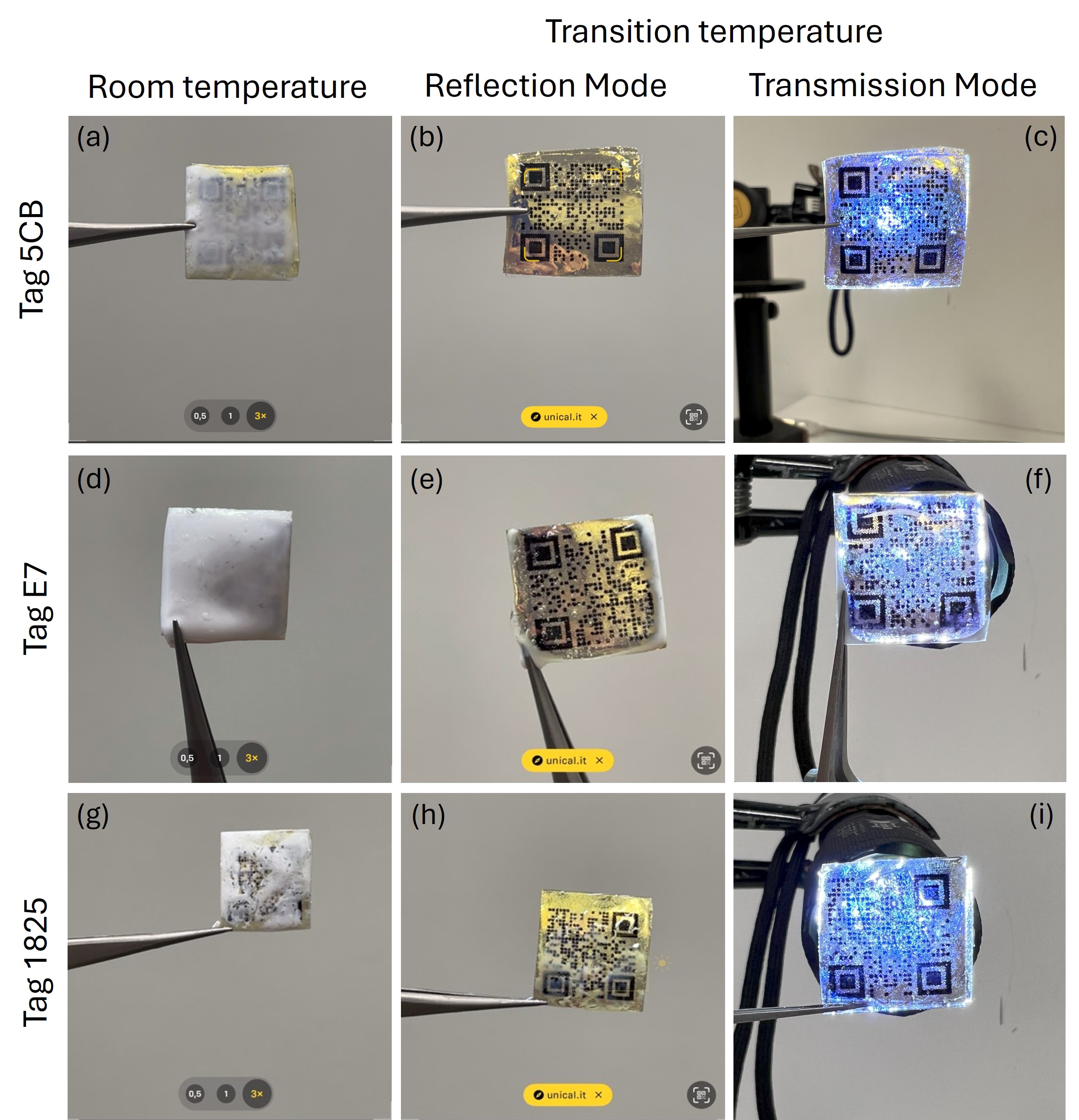}
    \caption{Snapshots of PUF with (a) TAG 5CB,  (d) TAG E7  and (g) TAG 1825 at room temperature where QR codes are not visible; (b) TAG 5CB, (e) TAG E7 and (h) TAG 1825 at transition temperature respectively with QR code reading, in reflection mode; (c) TAG 5CB, (f) TAG E7 and (i) TAG 1825 in transmission mode.}
    \label{img2}
    \end{figure}

\noindent The final tags have a thickness of a few hundred microns, see a schematic representation in Figure \ref{Img5}. 
The LCs used, including 5CB, mixed in 50 wt \%, E7, mixed in 30 wt \%, and 1825 mixed in 10 wt \%, were chosen due to their different properties, including birefringence and nematic-isotropic transition temperature. In the following, the name of the LCs are used to identify the tags.
At room temperature, the tags appear whitish, concealing both the QR code and the optical properties of the MIMI. Within seconds of heating the tags to their respective transition temperatures using a heat gun, the QR code becomes readable, and the optical properties of the MIMI are visible. This occurs because, upon  heating, the liquid crystals transition from the nematic to the isotropic mesophase, causing the mixtures to become transparent.
Figure \ref{img2} shows the smartphone camera snapshots of tags 5CB, E7 and 1825 at room temperature (Figure \ref{img2}a, \ref{img2}d and \ref{img2}g respectively) and at the respective transition temperatures, 35°C for tag 5CB (Figure \ref{img2}b), 59°C for tag E7 (Figure \ref{img2}e) and 140°C for tag 1825 (Figure \ref{img2}h). Once the LC transition point is exceeded, the QR code becomes easily readable, enabling access to the first security level. The exact transition temperature of the PDLCs were retrieved by heating the tags through a hot stage (see supporting videos S1, S2, S3).
This procedure also activates the second security level. However, this does not compromise the security efficiency of the tags, as the two levels are based on distinct challenges: QR code recognition and spectral identification, which depends on the optical operation mode — reflection or transmission. The gold color, resulting from the reflective properties of the nanocavity, is clearly visible. When the three tags are instead back-illuminated with a white light source, a blue color emerges, highlighting the transmission properties of the MIMI structure. (Figure \ref{img2}c,f, i).
    
\noindent The tag 5CB is designed for use in environments with temperatures below 35°C, ensuring that the QR code remains hidden at typical room temperatures. However, due to its low transition temperature, the nematic-isotropic phase change can be easily induced, making the QR code visible by simply rubbing the tag between hands for a few seconds (see supporting video S4). This tag offers a simple yet effective solution for anti-counterfeiting. For outdoor or high-temperature environments, the other two tags, E7 and 1825, with higher transition temperatures, provide suitable alternatives.

\begin{figure}[h]
    \centering
    \includegraphics[width=0.5\linewidth]{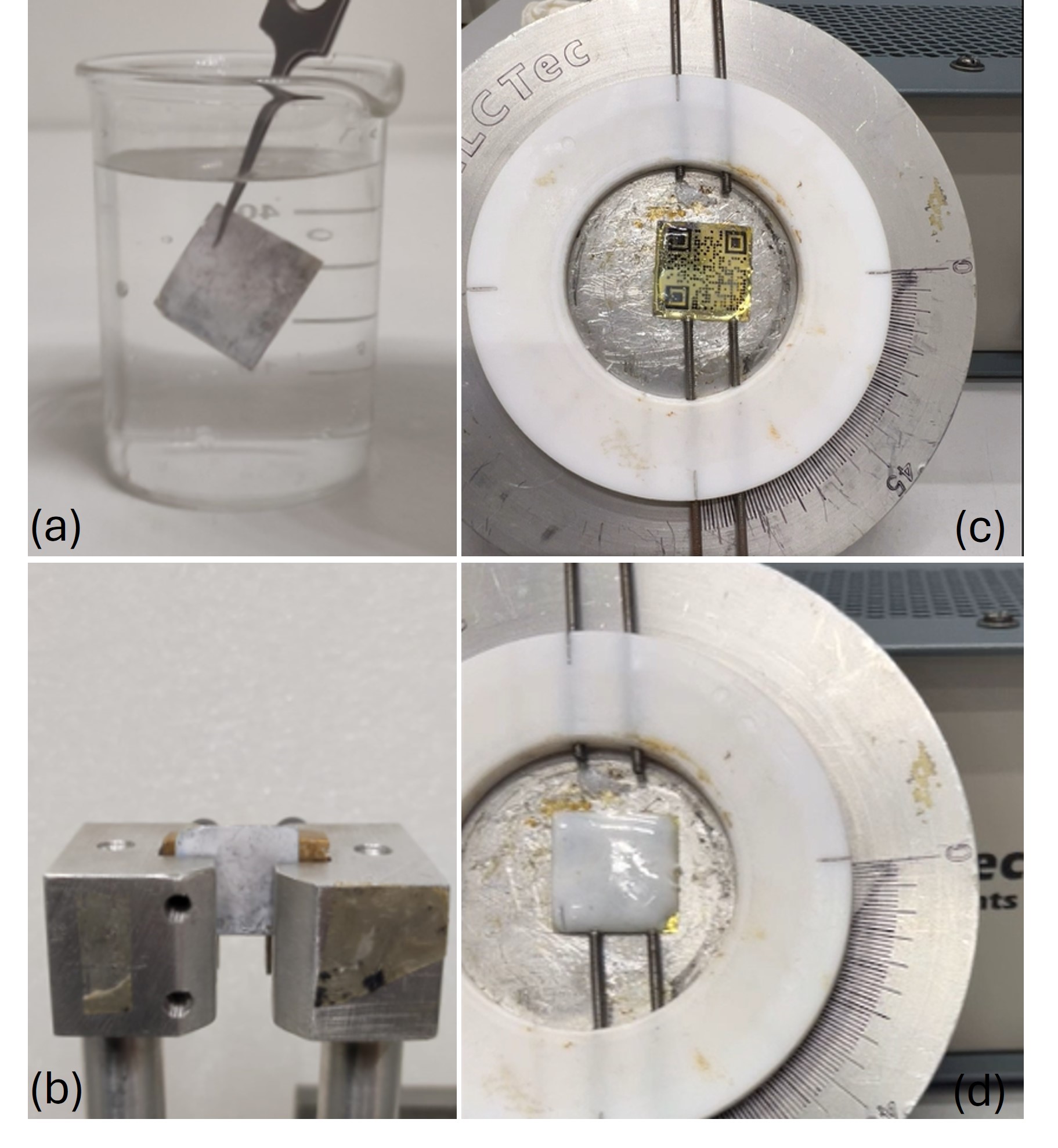}
    \caption{a) TAG E7 immersed in water. The process was repeated 20 times and integrity was verified. TG E7 on the mechanical elongation system, where the tag has been repeatedly stretched 30 times. c) TAG E7 on the hot stage. It was heated and subsequently d) cooled 50 times.}
    \label{img4}
\end{figure}

\noindent Several tests were conducted to assess the robustness of the realized tags. The first experiment evaluated the water resistance of the tags. As reported in (Figure \ref{img4}a), tag E7 was immersed in a beaker full of water for 30 minutes. Once the tag was removed from the water and heated, the QR code was fully readable. This procedure was repeated 20 times, and we can confirm that the properties of both the liquid crystals and PMMA were unaffected by water.
Next, a mechanical test was performed on the same tag using a custom-built setup where the tag was mechanically bent 100 times (Figure \ref{img4}b). Afterward, the tag was heated, and the QR code was successfully read. We can conclude that the properties of the tag remain intact even after mechanical stress. Finally, heating and cooling tests were conducted 50 times (Figure \ref{img4} c). The tag was placed in a laboratory hot stage, and the temperature was raised to the isotropic transition point of the liquid crystal. After reaching this temperature, it was cooled back to room temperature (Figure \ref{img4} d). After 50 cycles, the QR code remained easily readable.
Moreover, the structural color due to the nanocavity remained visible, indicating that the substrate was not affected by water immersion, the bending test, or the heating procedure.

\section{Conclusion}
The presented physical unclonable functions (PUF)-inspired system features two different security levels: QR code, camouflaged by PDLC, and the transmission and reflection colors enabled by the unique metamaterial nanocavity. At room temperature, the tags appear whitish due to the light diffusing behavior of the liquid crystals in the nematic phase. The QR code becomes readable within seconds by heating the tags to its liquid crystal transition temperature.  
Futhermore, the proposed solution addresses the issue of the QR code clonability by camouflaging it; it cannot be simply printed on paper. The QR code must be embedded in a PDLC, which also exhibits the optical properties of both the liquid crystal and the MIMI structure. This combination of materials and fabrication techniques does not affect the authentication process. Users can authenticate the tags without the need for expensive laboratory equipment - only a heat gun or hot stage and a smartphone are required to read the code. In certain environments, the tag 5CB can be even heated by rubbing it between the hands to reveal the QR code. These tags are easy to manufacture, cost-effective, and suitable for anti-counterfeiting applications across a wide range of products. The approach demonstrates the potential for dynamic authentication systems that can adapt to various environmental conditions without compromising security.
The water immersion, bending, and repeated heating and cooling tests confirm the durability and integrity of the tags over time.
Additionally, the versatility of these tags allows for the integration of further security levels. Future improvements, such as, the use of different liquid crystals or addition of fluorescent dyes, could enhance their security even more.

\bibliography{sample}

\end{document}